\documentclass[10pt, conference, compsocconf]{IEEEtran}
\PassOptionsToPackage{bookmarks={false}}{hyperref}
\usepackage{graphicx}
\usepackage{comment}
\usepackage{longtable}
\usepackage{verbatim}
\usepackage{setspace}
\usepackage[
  pdftitle = "{Requirements Analysis and Management for Benefiting Openness}",
    pdfauthor = "{Linaker and Wnuk}"
]{hyperref}
\usepackage[pdftex]{color}
\usepackage{multirow}
\usepackage{mathptmx}
\usepackage{array}
\usepackage{balance}
\usepackage{todonotes}

\begin{document}

\author{
\IEEEauthorblockN{Johan Lin{\aa}ker}
\IEEEauthorblockA{Department of Computer Science\\
Lund University, Lund, Sweden\\
Johan.Linaker@cs.lth.se}
\and
\IEEEauthorblockN{Krzysztof Wnuk}
\IEEEauthorblockA{Department of Software Engineering\\ 
Blekinge Institute of Technology, Karlskrona, Sweden\\
Krzysztof.Wnuk@bth.se}
}

\title{Requirements Analysis and Management for Benefiting Openness }

\maketitle
\thispagestyle{empty}

\begin{abstract}

Requirements Engineering has recently been greatly influenced by the way how firms use Open Source Software (OSS) and Software Ecosystems (SECOs) as a part of their product development and business models. 
This is further emphasized by the paradigm of Open Innovation, which highlights how firms should strive to use both internal and external resources to advance their internal innovation and technology capabilities.  
The evolution from market-driven requirements engineering and management processes, has reshaped the understanding of what a requirement is, and how it is documented and used. In this work, we suggest a model for analyzing and managing requirements that is designed in the context of OSS and SECOs, including the advances and challenges that it brings. The model clarifies how the main stages of requirements engineering and management processes can be adjusted to benefit from the openness that the new context offers. We believe that the model is a first step towards the inevitable adaptation of requirements engineering to an open and informal arena, where processes and collaboration are decentralized, transparency and governance are the key success factors.


\end{abstract}

\section{Introduction}




Software-intensive firms continuously have to face new challenges in order to sustain their competitive advantage. Known factors that they have learned to identify, and to a large degree control over the years include frequent technology changes~\cite{Aurum2007} and shifting market needs~\cite{Regnell2005}. During the past decade, the advent of various forms of openness, such as Software Ecosystems (SECOs)~\cite{jansen2013software} and Open Source Software (OSS) has become pivotal parts of many firms' product development and business models. This has pushed them into a new unknown context where they need to \textit{``learn how to play poker as well as chess''}~\cite{chesbrough2006open} by searching for, experimenting with, and ultimately using externally generated innovation. This new open context may be further explained by the Open Innovation (OI) model, which highlights how firms should strive to use both internal and external resources to advance their internal innovation and technology capabilities~\cite{Munir2015}.

 
Requirements Engineering (RE) has primarly focused on internal (from within the firm) stakeholder interaction in regards to activities such as analysis, research and development~\cite{WnukOIESEM2012}. 
Market-Driven Requirements Engineering~\cite{Regnell2005} and crowdsourcing~\cite{hosseini2014towards} brought more focus on external stakeholders. 
OI has not only blurred the boundaries between internal and external contexts, but also greatly extended the scope of requirements activities to entire ecosystems or open communities. Moreover, OI brought the support for both the adoption of externally acquired innovation and the active commercialization of internally generated innovations that are not aligned with the current business model (e.g. via licensing or sale)~\cite{chesbrough2006open}. As a result, and due to strong relationship between requirements engineering and value creation~\cite{Aurum2007}, requirements engineering for OI need to be reshaped to better \textit{``sustain innovation''}~\cite{Kauppinen}. As pointed out by Rohrbeck et al.~\cite{Rohrbeck20131593} value should no longer only be created for customers but must also be captured for partners and suppliers in increasingly more complex and collaborating ecosystems.  

In our previous work, we investigated OI in a large organization that recently transitioned to an OI model by abandoning the development of a purely proprietary code base for their software product and making use of an OSS project (referred as a \textit{"platform"}) as a source of innovation (in both knowledge and technology)~\cite{WnukOIESEM2012}. The platform is not only the main component of the firm's product, but is also available to, or used by any other player within the SECO. The focus on previous investigation was on identification of requirements management~\cite{hood2007requirements} and decision making challenges associated with the adoption of OI.


In this vision paper, we propose a model for managing requirements that helps to benefit from openness and responds to previously outlined challenges. In particular, we believe that the model contributes in requirements elicitation, analysis, release planning and commercialization of internally generated requirements that are not aligned with the current products or business models.  

This paper is structured as follows. Section~\ref{sec:background} brings background and related work while Section~\ref{sec:RAMBO} outlines and explains our model. We conclude the paper in Section~\ref{sec:Concl} and discuss future work direction. 

\section{Background}\label{sec:background}
Here we present how OI connects to Software Engineering, and how RE differentiates in an open versus a closed context.


\subsection{Software Engineering in Open Innovation}
\begin{figure}
\begin{center}
\includegraphics[width=\columnwidth]{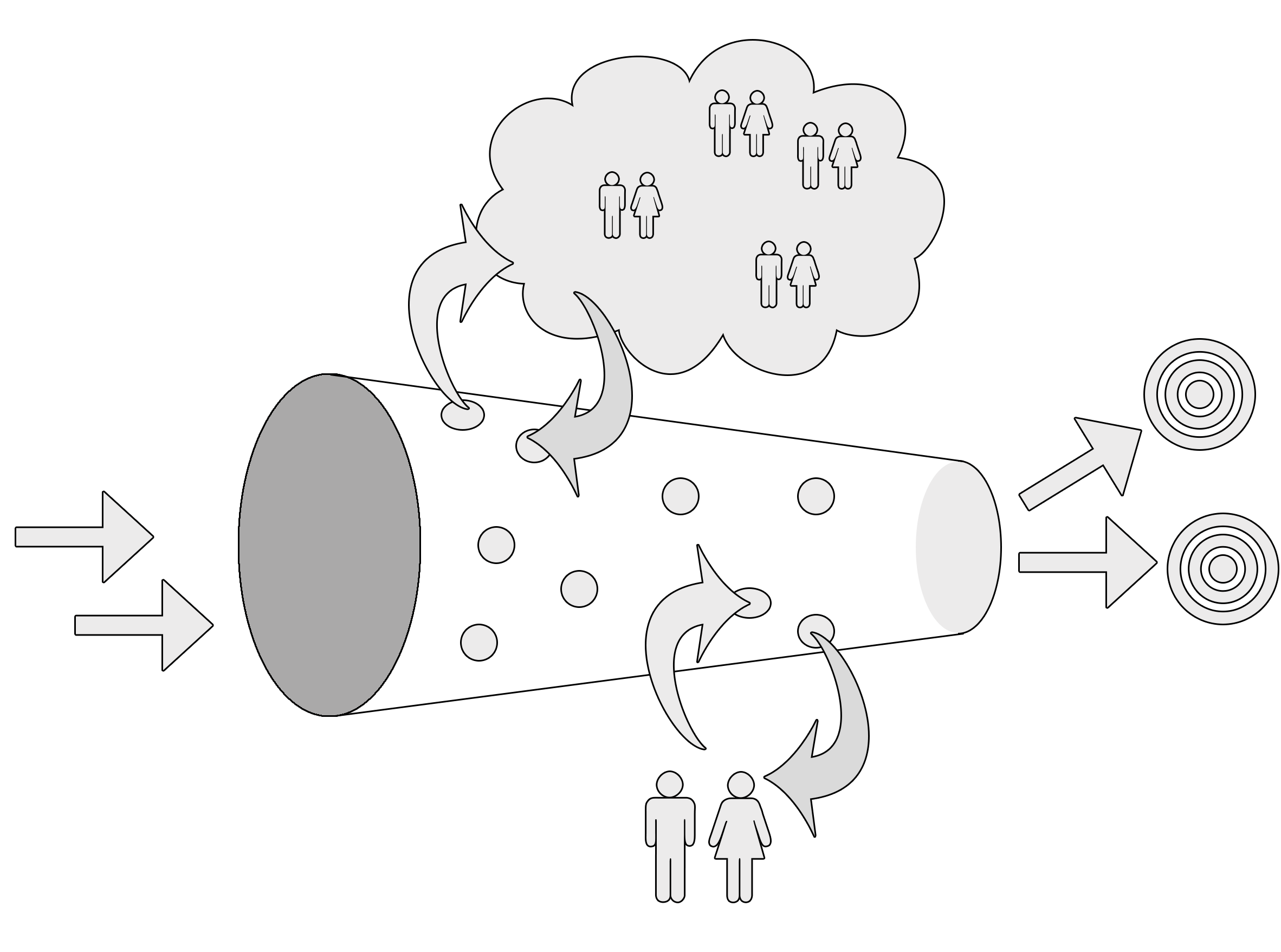}
\caption{The OI model illustrated with interactions between the firm (funnel) and its external collaborations. Adopted from Chesbrough~\cite{chesbrough2006open}.}
\label{fig:Funnel}
\end{center}
\end{figure}

The OI model is commonly described with the use of a funnel~\cite{chesbrough2006open}, see Fig.~\ref{fig:Funnel}. The funnel itself represents the software development process. This may in turn be illustrated either as an iterative process, or just as one single development cycle (e.g. from RE, design, implementation, test, to release). The funnel is full of holes representing openings between the firm's development process and the open environment, in our case, an OSS ecosystem. The arrows going in and out represent transactions and exchange of knowledge between the two, such as feature requests or bug reports, design proposals or feature road maps, feature implementations or bug fixes, test cases, complete sets of code as in plugins, components, platforms or even products. From the RE perspective this includes activities such as elicitation, prioritization and release planning, which are performed interactively between the internal and the open environment.

The knowledge exchange illustrated in Figure~\ref{fig:Funnel} may be bi-directional, i.e. it can go into the development process from the open environment (\textit{outside-in}), or from the development process out to the open environment (\textit{inside-out}). On the left-hand side the arrows represent the intent and business model, which encompasses and motivates the development process. On the right-hand side the arrows represent the output from the development process, with some degree of innovation. This output may be a new or improved product or service from which the company may capture or create value with the help of their business model. Also on the right-hand side are the bulls-eye marks, which represents the current but also new alternative markets to which the firm's business model could deliver the product or service produced from their development process.

 
\subsection{Requirements Engineering in Open Innovation} 
\begin{figure*}
\begin{center}
\includegraphics[width=\textwidth]{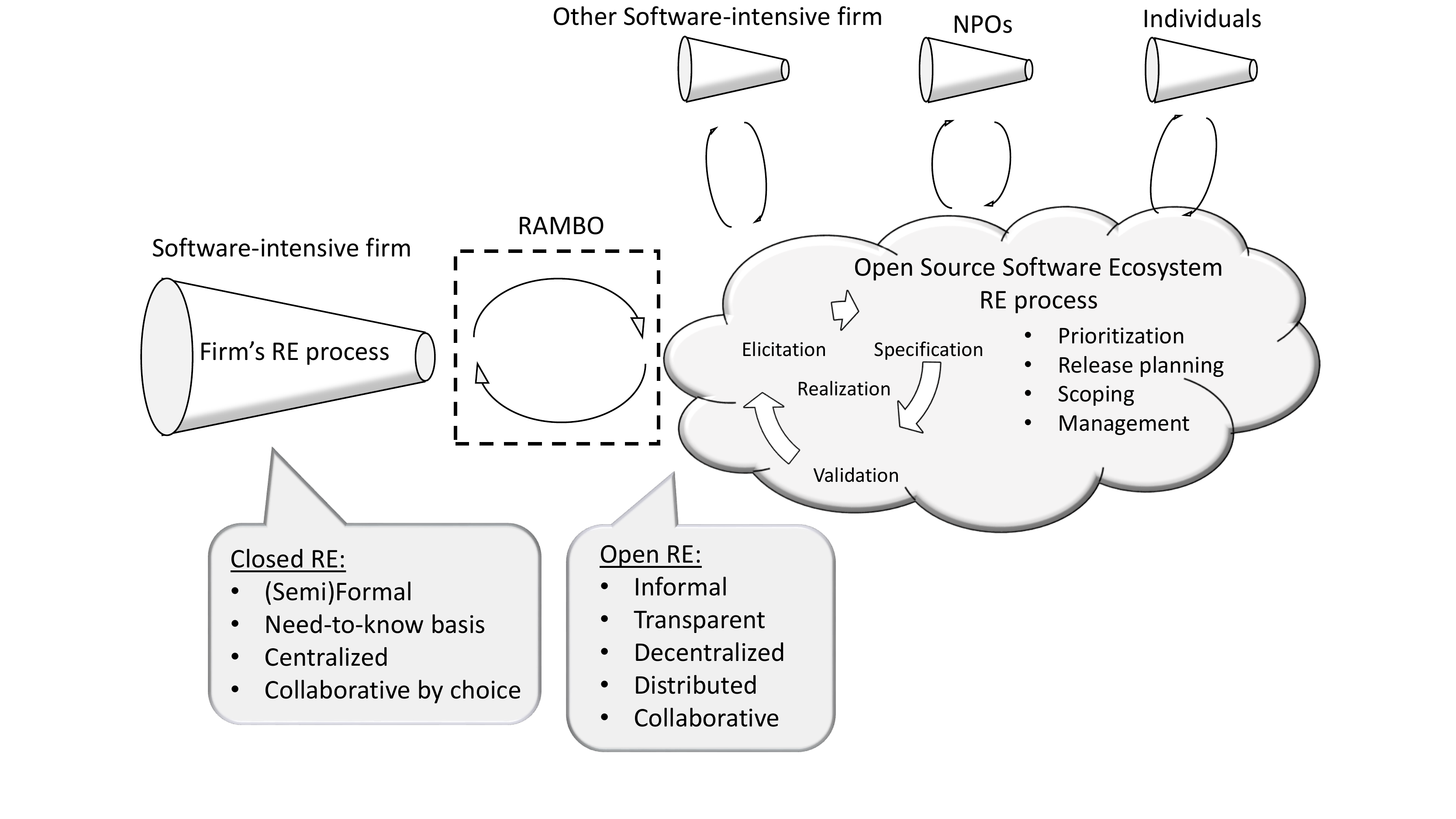}
\vspace{-1cm}
\caption{The requirements engineering context influenced by the OI paradigm. }
\label{fig:REinOI}
\vspace{-0.5cm}
\end{center}
\end{figure*}

Fig.~\ref{fig:REinOI} presents RE in the context of OI. An example software-intensive firm is represented by a funnel, connecting to the OI model. The open environment, or a OSS ecosystem, with which the firm interacts, is represented by the cloud. Other actors, which for example may include individuals, NPOs and other firms are also represented by a funnel each. We further need to differentiate between an internal and external RE process. With internal, we refer to the process connected to a firm's internal software product development. The external concerns the one used in an OSS ecosystem with which the firm, and all other actors, interact with. 




Looking at OSS ecosystems, requirements practices are often informal and overlapping~\cite{scacchi_understanding_2009-1}. Requirements are commonly asserted through transparent discussions and suggestions by the OSS project's developers and users, often together with prototypes or proof-of-concepts~\cite{alspaugh_ongoing_2013, german_gnome_2003}. Assertion may also be done post-hoc, simultaneously as the requirement realization~\cite{ernst_case_2012, german_gnome_2003}. These assertions are specified and managed in what Scacchi refers to as informalisms~\cite{scacchi_understanding_2009-1}, e.g. reports in an issue tracker, messages in a mailing list, or commits in a version control system. Through social interaction facilitated by the infrastructure persisting the informalisms, requirements are further enriched and validated~\cite{torkar_adopting_2011-1, ernst_case_2012, german_gnome_2003}. Prioritization is commonly conducted by ecosystem maintainers overseeing the project management, though care is often taken to the opinions of other developers and users (e.g. through votes and comments~\cite{laurent_lessons_2009}). Ernst \& Murphy refers to this lightweight and evolutionary process of requirements refinement as Just-In-Time (JIT) requirements (illustrated by the circular arrows inside the OSS ecosystem in Fig.~\ref{fig:REinOI}), compared to the more traditional upfront requirements characterized by heavy processes and tool support~\cite{ernst_case_2012}. Further, Alspaugh \& Scacchi contrasts how OSS RE steps away from what they refer to as Classical Requirements, characterized as having a central repository, with requirements defined in the problem space, describing the product of need, along with processes for examining the requirements for completeness and consistency~\cite{alspaugh_ongoing_2013}. For better consistency, we choose to re-label JIT and OSS RE as Open RE, and that described as traditional upfront, and classical requirements as Closed RE.

Release-planning in OSS ecosystems is often employed using either a feature-based and time-based strategy~\cite{michlmayr_release_2007}. The feature-based performs a release when a set goals have been fulfilled, e.g. a certain set of features has been implemented. The time-based performs releases according to a preset schedule. Hybrid versions has also used~\cite{wright_subversion_2009}. Earlier work has reported a favor towards the time-based strategy~\cite{michlmayr_why_2015,wright_subversion_2009}. With fixed release dates and intervals, firms can better adapt their internal plans so that additional patchwork and differentiating features may be added in time for product shipment to market~\cite{michlmayr_why_2015,michlmayr_release_2007}. Other issues associated with time-based release strategy include: rushed implementations, workload accumulation, outdated software and delayed releases~\cite{michlmayr_why_2015}.

Open RE, compared to Closed RE, can be seen as being informal to different degrees, e.g., to what level requirements are analyzed and managed~\cite{ernst_case_2012}. Requirements are often decentralized and distributed over multiple sources, often with a limited tracing. Influence and participation in the work and decision-making are also distributed. Discussions and steering documents are all public and transparent for anyone to see, or participate. Collaboration and negotiation about requirements are key, as consensus often is needed to make certain decisions~\cite{aagerfalk_outsourcing_2008}. These distinctions between Open and Closed RE visualize a process and knowledge gap that commonly exist between firms and OSS ecosystems on the context of OI~\cite{Munir2015, WnukOIESEM2012}. This gap need to be addressed to the firms can increase their benefits from ecosystem participation (both monetary and non-monetary), ecosystems can get more active players and enriched discussions and influence and customer can receive better products with shorter time to market and greater degree of innovation. This is what we aim to address with the construction of the Requirements Analysis and Management for Benefiting Openness (RAMBO) model. The model focuses on the interaction and overlap between the internal RE process of the focal firm, with that of its connected OSS ecosystem, to better manage the challenges implied by OI.

\section{Requirements Analysis and Management for
Benefiting Openness (RAMBO)}\label{sec:RAMBO}


\begin{figure*}
\begin{center}
\includegraphics[width=\textwidth]{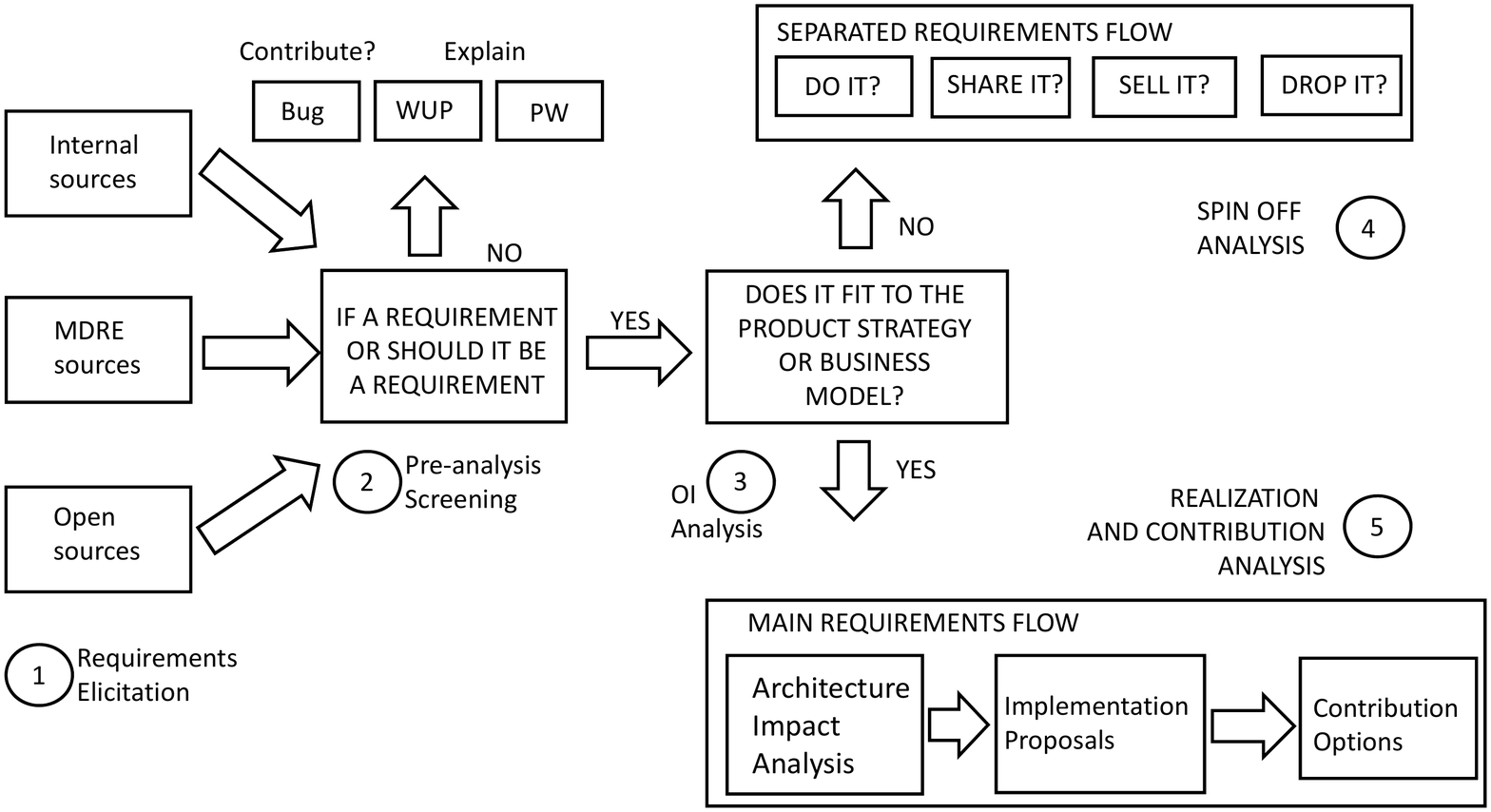}
\vspace{-2cm}
\caption{The Requirements Analysis and Management for Benefiting Openness (RAMBO) model. }
\label{RAMBO}
\vspace{-0.5cm}
\end{center}
\end{figure*}


Figure~\ref{RAMBO} outlines the RAMBO model. The model is divided into five phases outlined in the subsections that follow: 1) requirements elicitation, 2) requirements screening and pre-analysis, 3) Open Innovation potential analysis, 4) spin off analysis and 5) realization and contribution analysis.  
In the context of this work we follow the definition of Requirements Management (RM) provided by Hood et al. as ``the set of procedures that support the
development of requirements including planning, traceability, impact
analysis, change management and so on... and the sum of the interfaces between
requirements development and all other systems engineering disciplines
such as configuration management and project management.''~\cite{hood2007requirements}. We also recognize the role of requirements management in supporting managing requirements between products and projects.  

\subsection{Step 1: Elicitation}

Requirements Elicitation in OI (marked with 1 in Figure~\ref{RAMBO}) differs greatly from previously known and published contexts. During requirements elicitation, a firm that operates in OI needs to quickly and properly identify which of the sources of requirements to use, and for what purpose. OI still contains the ''classical`` internal and MDRE requirements sources, but is extended by the open sources of requirements.

The internal requirements sources (typical for bespoke RE) are characterized by a small and well known set of customers or stakeholders whose wishes or constraints are important and taken into consideration. Apart from that, any firm that sells a product on an open market needs to elicit requirements from the MDRE sources through a balance of market-push and technology-pull~\cite{Regnell2005}. Wishes from customers, both known and unknown, are elicited through more traditional methods (e.g., interviews) complemented with approaches such as market analysis. New functionality and innovative requirements are added internally as R\&D progresses and the product evolves. 

Regarding open sources, such as an OSS ecosystem, requirements are both market- and technology-driven as in MDRE. A difference is that in an OSS ecosystem, it usually is the end-users who are actively suggesting new requirements as well as driving the innovation forward by co-developing the products. However, active elicitation may still be needed, but through new approaches such as participation in discussions facilitated by the ecosystem infrastructure (e.g., mailing-lists, issue-trackers and IRC-channels) and at off-line events (e.g., user-conferences and hackathons). Crowdsourcing as a requirements elicitation technique is also highly recommended here as it increases the quality
and comprehensiveness and even the economic feasibility of requirements elicitation~\cite{hosseini2014towards}. Moreover, requirements elicited from the crowd rather accurately represent the needs and pains of that crowd, allowing to skip a large part of requirements validation process and providing quick feedback. Stakeholder discovery is also greatly supported by crowdsourcing that provides opportunities to create user personality clusters~\cite{Groen2015} and filter their opinions about the products and requested requirements. This greatly supports user feedback analysis and helps in defining new requirements.
Finally, the identification, analysis and prioritization of present stakeholders is key to find those relevant, and can bring valuable requirements that increase profitability. 


Embracing and supporting creativity is important as this stage as new ideas that both fit into the current product portfolio and are way beyond it are much appreciated. The illumination and verification parts of creativity workshops can greatly support requirements discovery and idea generation that can be further detailed in OI contexts~\cite{Maiden1}. Automated support for creativity is highly desirable here as the number of unfamiliar connections between familiar possibilities of requirements is high and can overload requirements analysts~\cite{Bhowmik}.	

As with MDRE, the number of requirements grows fast in an OSS ecosystem, creating potential problems in regard to managing quality aspects such as dependencies and complexity. Another aspect is that prioritization and throughput of requirements suffers due to the large repositories. Specific for the OI context however, is that the requirements are spread out in several decentralized repositories (informalisms~\cite{scacchi_understanding_2009-1}), often very unstructured and expressed both through a variety of implementations and natural language descriptions.

\subsection{Step 2: Pre-Screening and Analysis}


After all relevant sources (Open, MDRE and internal if relevant) are identified and prioritized, the requirements originating from these sources need to be screened. Requirements screening in OI is significantly challenging due to the following reasons: 1) they are spread out in several decentralized repositories, 2) they are unstructured and 3) everyone has access to them and can already be working on their implementation, 4) consideration often has to be taken to others' opinion (e.g., in a meritocracy), and 5) many of the requirements are actually bug fixed, Wrong Usage of a Product (WUP) (user of customer misunderstood how to use the product) or Problems that have Workarounds (PW) and can be solved without additional implementations. 

Therefore, the main part of the screening process is the decision if the analyzed piece of information is a requirements or should be a requirement. Here, we would like to stress that RE in OI should adopt a broad definition of what a requirement is, which also includes needs for stakeholders that are not relevant for a given product but could be relevant for a spin-off product. This implies that the product boundaries or constraints are greatly removed and make a vast majority of requirements potentially interesting or relevant. Therefore, RE in OI requires a lightweight and efficient method for quickly investigating the relevance and novelty of potential requirements, along with the potential cost and contribution strategy in consideration. Previous work by Maalej and Nabil~\cite{Maalej} provides promising results for classifying app reviews as bug reports, feature requests, user experiences and ratings. Laurent et al. ~\cite{Laurent} proposed a method for automated requirements triage that cluster incoming stakeholder requests into hierarchical feature sets. We argue that their work should be extended by adding OI-specific concerns, e.g. contribution potential and realization options.

The challenge here remains in analyzing incoming requirements with openness in mind, e.g. taking the broader perspective than suggested by Davis effort, dependency and important analysis, multiple release planning~\cite{Davis2003} or impacted business goals suggested by Laurent et al. ~\cite{Laurent}. In this broader perspective, a requirements analysis should calculate the probability of completion if someone else from the ecosystem contributes in co-development or analyze optimistic or pessimistic scenarios with or without other contributions within the same OSS ecosystem. 
 
\subsection{Step 3: OI Analysis}


When relevant requirements are identified and screened, their OI potential need to be analyzed. 
This third step of the RAMBO model represents a significant difference between MDRE and RE in OI. 
In MDRE, requirements are analyzed through the lens of the current product portfolio, along with the stakeholders, their needs and future plans. 
Incremental innovation has higher changes to be selected as radical ideas or ideas not associated with significant stakeholders are considered risky or unprofitable. This was considered as a challenge in our previous investigations and resulted in many promising ideas been rejected~\cite{WnukOIESEM2012}. RE in OI needs to take a broader perspective and consider these ``not fitting'' requirements as potential spin-off ideas.
This addressed the well-known "PARC Problem" experienced by Xerox~\cite{chesbrough2006open}; the inability to assess and capture value for (technology) innovations that were not directly related to Xerox products. As quite often not all requirements can be implemented in the next release due to low priorities and lack of resources, the potential waste of unimplemented ideas can be shared with other ecosystem players. Therefore, step 3 of the model investigates if the potential requirements fit into the current product portfolio or future plans, or not. Even if some requirements do not fit into the current strategies, they should not stay ``locked'' internally in the requirements database or be down-prioritized by more urgent requirements, but made open and shared with other players who may find them more relevant for their offerings. 

\subsection{Step 4: Spin Off Analysis}  
Requirements that are considered as not the best fit for the current product portfolio, can and should be shared with other ecosystem players. Moreover, potential spin-offs should be discussed and executed to benefit from potential novelty and innovation included in those requirements. This step often involves making alliances with other ecosystem players or spinning-off start-up companies that can develop and monetize these ideas. The important factor in this step is also to make an effort on keeping the key personnel away from leaving the firm together with the new ideas. This step corresponds to the inside-out part of the knowledge exchange in the OI model, see Figure~\ref{fig:Funnel}. Open requirements and increased transparency in release plans are important at this stage. For example, assuming that two ecosystem players are working jointly on a feature that is going to provide substantial benefits for the customers, there should to be a discussion and agreement about release plans synchronization so that both players can benefit from the feature. These release plans should also be synchronized with previously agreed contribution strategy for the feature so that maintenance costs can be directly minimized via commoditization. Finally, ideas may also be dropped if considered not interesting enough or economically viable. 

\subsection{Step 5: Realization and Contribution Analysis} 

Requirements considered as a good match for the current product strategy need to be analyzed from the realization and contribution perspectives. First, the Architecture Impact Analysis (AIA) need to be performed to understand the impact of new requirements on the current architecture and potential changes that need to be made during the implementation. The involvement of software architect as a technical counterpart of software product manager is important at this stage to ensure the alignment between product requirements and architecture and create a stepwise architectural evolution that fits with the product roadmaps~\cite{Lucassen2015}.

Second, the Implementation Possibilities (IP) (adaption of OSS solution, differentiation based on commodity offered by OSS, or fully own implementation) are analyzed and considered. Each of the three options need to be analyzed from the potential value and long-term cost perspective. The risk associated with fully own implementation is a significant change to the common code base and several patches. Moreover, the OSS ecosystem may in the near future come up with a solution to exactly the same requirement that may be equally as good, or even better than the firm's own.  Finally, Contribution Options (CO) need to be considered, and the scope and time line of contributions need to be set. Both realization and analysis of contribution options require the following two elements: 1) a good understanding of the value from both customer and internal business perspectives~\cite{SMR1560} and 2) an understanding of the commoditization process for a given product or market that will help in estimating if and when a given implementation should be contributed to the open community.~\cite{van2009commodification}.  


\section{Conclusions and Future Work}\label{sec:Concl}


In this paper, we present our vision of the Requirements Analysis and Management Model for Benefiting Openness that we designed with the specific challenges in mind that Open Innovation brings to Requirements Engineering. Our model recognizes the challenging aspects of stakeholder and requirements elicitation, triage, analysis and implementation that Open Innovation brings. We have summarized the RAMBO model in five steps and explain how each of the steps differ from commonly accepted requirements engineering practice. 

Our vision need significant further work that we plan in the near future. Firstly, we plan to study each of the associated requirements phases in more detail and bring more evidence of the lack of support for these activities. We plan to use our industry network to search for empirical evidence if the activities outlined in the RAMBO model are currently performed and in what way or maybe not performed. Secondly, we plan to select one large company that operated in OI and study its requirements process to create process improvement action plan that will encourage them to adapt the parts of the RAMBO model that are currently not present. Finally, we plan to create a governance model for RE in OI that will contain all relevant activities and decision points that a software-intensive firm should consider to better utilize the benefits that openness offers.



\bibliographystyle{plain}
\bibliography{RAMBO}
\end{document}